# Reverse circling supercurrents along a superconducting ring

Tian De Cao

*Department of Physics, Nanjing University of Information Science & Technology, Nanjing 210044, China*

The reason why high temperature superconductivity has been being debated is that many basic ideas in literatures are wrong. This work shows that the magnetic flux quantum in a superconducting ring has been inaccurately explained in fact, thus we suggest a reinterpretation of the magnetic flux quantum in a superconducting ring on the basis of the translations of pairs. We also predict that the internal and external surface of a superconducting tube have the reverse circling supercurrents. This means that a more thick tube could trap a larger amount of flux. Both the magnetic flux quantum and the reverse circling supercurrents could not be found with the London equation.

London presented the magnetic flux quantum in a twofold-connected superconducting body (ring or tube) [1], the magnetic flux quantum has been confirmed with some low-temperature superconductors [2, 3] and with some high-temperature superconductors [4] under some discrepancy. However, we find the magnetic flux quantum in a superconducting ring have been inaccurately explained in fact.

Firstly, let us see the quantum mechanics explanation of the magnetic flux quantum. One assumes that the superconducting pairs have the wave function $\psi(\vec{x}) = \sqrt{\rho(\vec{x})} e^{i\theta(\vec{x})}$, and this gives the supercurrent density

$$\vec{j}(\vec{x}) = -\frac{2e\hbar}{m_e} \rho \vec{\nabla} \theta(\vec{x}) + \frac{4e^2}{m_e} \rho \vec{A}(\vec{x}) \qquad (1)$$

When $\vec{j}_s = 0$ in the superconductor, they find $\hbar \vec{\nabla} \theta - 2e\vec{A} = 0$ and conclude $\oint_L \vec{A} \cdot d\vec{l} = \frac{1}{2e} \hbar \oint \vec{\nabla} \theta \cdot d\vec{l} = \Delta\theta = \pm \frac{1}{e} \hbar n\pi = \pm \frac{h}{2e} n = \pm n\phi_0$. However, they have not given the concrete expression of the phase function $\theta(\vec{x})$, thus they hove not confirmed $\oint_L d\vec{l} \cdot \vec{\nabla} \theta(\vec{x}) = \Delta\theta = \pm 2n\pi$. Someone may argue that $\psi(\vec{x},t) = \sqrt{\rho(\vec{x},t)} e^{i[\theta(\vec{x},t) + 2n\pi]} = \sqrt{\rho(\vec{x},t)} e^{i\theta(\vec{x},t)}$ would give $\Delta\theta = 2n\pi$, this is obviously an arbitrary conclusion. Secondly, the second GL equation [5] also gives the supercurrent equation similar to Eq.(1). Gorkov derived the GL equation from the BCS theory [6]. Moreover, some strongly correlated models have also given similar equations. However, as discussed above, they did not find the concrete form of $\theta(\vec{x})$, thus the magnetic flux quantum has not been explained in fact.

We find that the superconducting energy gap necessarily lead to the disappearance of some quasi-electrons [7], thus the superconducting pairs must behave like bosons (however, it is suggested that the bosons should be observed on the related theories because the bosons could not be observed by some experiments), this means that a boson-fermion model is necessary for describing some properties of superconductors. This behavior is different from the non-Fermi liquid behavior of low dimension electron systems in which the fermions number is a conservation amount. On this boson-fermion model we derived a new supercurrent equation. Because the parameters $C\alpha$ and $C\beta$ in the previous article can be determined as $C\alpha = \dfrac{\hbar}{2m_e}$ and $C\beta = \dfrac{2}{m_e}$ by the dimensional analysis, we find the supercurrent density

$$\vec{j}(\vec{x}) = -\dfrac{e\hbar}{m_e} n_s \vec{\nabla}\theta(\vec{x}) - \dfrac{2e^2}{m_e} n_s \int \theta(\vec{x}-\vec{x}')\vec{A}(\vec{x}')d^3 x' \tag{2}$$

where we have expressed $\sum_{\vec{q}} n_B(\Omega_{\vec{q}}) = n_s$ and defined the function

$$\theta(\vec{x}) = \dfrac{1}{n_s}\int d^3 q\, n_B(\Omega_q) e^{i\vec{q}\cdot\vec{x}} \tag{3}$$

The Bose-Einstein distribution is $n_B(\Omega_{\vec{q}}) = 1/[e^{(\Omega_{\vec{q}}-2\mu)k_B T} - 1]$, $2\mu$ is the chemical potential of the boson systems while $\mu$ is the chemical potential of normal electrons, and the chemical potential is determined by the total electron number (the two times of boson number + the non-superconducting electron number = the total electron number). The Meissner effect can be explained, that is, $\vec{j}_s = 0$ in the enough deep position of a superconductor. Is $\theta(\vec{x})$ in Eq.(3) the phase function? This has to be answered, but we find $\oint_L d\vec{l}\cdot\vec{\nabla}\theta(\vec{x}) = \Delta\theta = 0$ on the basis of Eq.(3). Generally, the relation $\oint_L d\vec{l}\cdot\vec{\nabla}\theta(\vec{x}) = \pm 2n\pi$ requires $\theta(\vec{x}) = \vec{p}\cdot\vec{x}/\hbar + \vartheta(\vec{x})$, $\vartheta(\vec{x})$ is zero or not and $\vec{p}$ is the canonical momentum. In fact, no one has derived the so-called phase function which gives $\Delta\theta = 2n\pi$ with $n\neq 0$.

The phase function relating to the canonical momentum, $\theta = (\vec{p}-e\vec{A})\cdot\vec{x}/\hbar$, has been used to explain the AB effect [8]. To explain the magnetic flux quantum, we suggest the ideas below. The moving pairs look like bosons, the plane wave function of the bosons is $\psi(\vec{x},t) = Ce^{i\varphi(\vec{x})}$ with $\varphi(\vec{x})=(\vec{p}-2e\vec{A})\cdot\vec{x}/\hbar$. Because $\vec{j}_s=0$ in the superconducting ring with and without the magnetic field, the integral along the ring gives

$\oint_L (\vec{p}-2e\vec{A})\cdot d\vec{l} = (n_1+\frac{1}{2})h$ and $\oint_L \vec{p}\cdot d\vec{l} = (n_2+\frac{1}{2})h$ (destructive interference), thus $\oint_L \vec{A}\cdot d\vec{l} = (n_2-n_1)\frac{h}{2e} = n\frac{h}{2e} = n\phi_0$. An exact explanation of the magnetic flux quantum should depend on the path integral approach, which is not discussed in this work. The particles talked about the AB effect are electrons, while the particles in a superconducting ring are bosons (superconducting pairs).

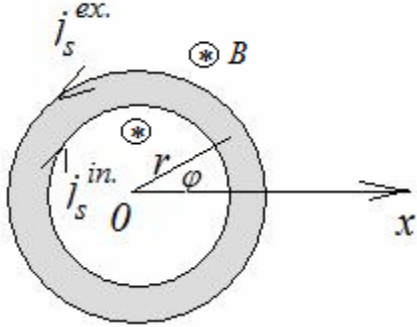

Fig.1: superconducting tube

Now let us consider a superconducting tube (or ring). The origin of coordinates of the position vector $\vec{x}$ in Eq.(3) is in a superconductor. If the origin of coordinates is taken at the geometric center of a superconducting tube as shown in Fig.1, we should do the transition $\vec{x} \to \vec{x}-\vec{x}_0$ for the function $\theta(\vec{x})$. When $\vec{j}(\vec{x})=0$ in the tube which is enough thick one, we have

$$\hbar\vec{\nabla}\theta(\vec{x}-\vec{x}_0) + 2e\int \theta(\vec{x}-\vec{x}')\vec{A}(\vec{x}')d^3x' = 0 \qquad (4)$$

The integration along the tube gives $\hbar\oint_L d\vec{x}\cdot\vec{\nabla}\theta(\vec{x}-\vec{x}_0) + 2e\oint_L d\vec{x}\cdot[\int \theta(\vec{x}-\vec{x}')\vec{A}(\vec{x}')d^3x'] = 0$. Because $\oint_L d\vec{x}\cdot\vec{\nabla}\theta(\vec{x}-\vec{x}_0)=0$ no matter whether the supercurrent $\vec{j}=0$ or $\neq 0$, we obtain $\oint_L d\vec{x}\cdot[\int \theta(\vec{x}-\vec{x}')\vec{A}(\vec{x}')d^3x'] = 0$ for the position vectors $\vec{x}$ of supercurrent $\vec{j}=0$. Noting the symmetry of the tube, we use the cylindrical seat table, $\vec{x}=(r,\varphi,z)$, we find the mean value theorem for integrals can be applied to $\int \theta(\vec{x}-\vec{x}')\frac{r}{\tilde{r}}\vec{A}(\vec{x}')d^3x'$ and we obtain $\oint_{\tilde{L}} \vec{A}(\tilde{\vec{x}})\cdot d\tilde{\vec{x}} = 0$ for the position vectors $\tilde{\vec{x}}$ of supercurrent $\vec{j}\neq 0$ because $\tilde{\vec{x}} \neq \vec{x}$ should be always met, where $\tilde{\vec{x}} = (\tilde{r}, \tilde{\varphi}, \tilde{z})$ while $d\tilde{\vec{x}} = \tilde{r} d\tilde{\varphi} \vec{e}_\varphi$. Strictly speaking, if the change of $A_\varphi(r,\varphi,z)$ along with $r$ is monotonic, the mean value theorem for integrals can be used, thus we have $\oint_{\tilde{L}} \vec{A}(\tilde{\vec{x}})\cdot d\tilde{\vec{x}} = 0$; if the changes of $A_\varphi(r,\varphi,z)$ along with $r$ may be from positive to negative one, we also have $\oint_{\tilde{L}} \vec{A}(\tilde{\vec{x}})\cdot d\tilde{\vec{x}} = 0$. That $\vec{A}(\vec{x})$ do not depend on $\varphi$ is used. Because the critical supercurrent density in a superconducting tube is a finite value, and the reverse circling supercurrents intend to decrease the amount of flux in a tube ( or ring), thus a more thick tube could trap a larger amount of flux (the effect of the current in the external surface is weakened).

This result can not be derived from the London equation. The London equation is

$$\vec{j}(\vec{x}) = -\frac{e\hbar}{m_e} n_s \vec{\nabla}\theta(\vec{x}) - \frac{2e^2}{m_e} n_s \vec{A}(\vec{x}) \tag{5}$$

Because a general result is $\oint_L d\vec{l}\cdot\vec{\nabla}\theta(\vec{x}) = \Delta\theta = 0$, thus one should obtain $\oint_L \vec{A}(\vec{x})\cdot d\vec{l} = 0$ for $\vec{j}(\vec{x}) = 0$. This shows that both the magnetic flux quantum and the reverse circling supercurrents could not be found with the London equation.

The results are summarized as follows: the magnetic flux quantum $\oint \vec{A}(\vec{x})\cdot d\vec{x} = \pm n\phi_0$, $n\neq 0$, is for the position vectors $\vec{x}$ of supercurrent $\vec{j} = 0$ while the magnetic flux $\oint_{\tilde{L}} \vec{A}(\tilde{\vec{x}})\cdot d\tilde{\vec{x}} = 0$ is for the position vectors $\tilde{\vec{x}}$ of supercurrent $\vec{j}\neq 0$. This requires that the integral path $\tilde{L}$ is near the outer surface of the tube. This obviously shows that there should be the reverse circling supercurrents along the internal and external surface of a superconducting tube (or ring). Both the magnetic flux quantum and the reverse circling supercurrent are derived from a premise, that is, there is the solution of $\vec{j}=0$ at some position vectors $\vec{x}$ (while $\vec{j} \neq 0$ on the surface of the tube). The reverse circling supercurrents are easily examined by comparing the direction of the magnetic field outside and inside the tube. For example, the magnetic field directions near the tube are shown in Fig.1. If this prediction is confirmed, the supercurrents in superconducting tubes should be larger than the ones suggested in previous experiments, and this should be also noted by physicists. Other wrong ideas on superconductivity will be argued with other evidences.